\documentclass[twocolumn,prb,showpacs]{revtex4}

\usepackage{amsmath}
\usepackage{amssymb}
\usepackage{epsfig}

\newcommand{\tr}{\text{Tr}}
\def\UnitMatrix{\mbox{1\hspace{-.25em}I}}

\begin{document}
\title{Quantum discord and quantum phase transition in spin chains}

\author{Raoul Dillenschneider}
\email[E-mail address : ]{raoul.dillenschneider@physik.uni-augsburg.de}
\affiliation{Department of Physics, University of Augsburg, D-86135 Augsburg, 
Germany}

\date{\today}

\begin{abstract}
Quantum phase transitions of the transverse Ising and 
antiferromagnetic XXZ spin $S=1/2$ chains
are studied using \emph{quantum discord}. Quantum discord allows
the measure of quantum correlations present in many-body quantum
systems. It is shown that the \emph{amount} of quantum correlations
increases close to the critical points. The observations are in agreement 
with the information provided by the concurrence which measures the 
entanglement of the many-body system.
\end{abstract}


\maketitle

\section{Introduction \label{Section1}}

Quantum phase transitions (QPT) occur when the ground state
of a many-body system at absolute zero temperature
undergoes a qualitative change by variation of
a coupling and/or an external parameter \cite{Sachdev}.
QPT are present in spin systems
as for example in the one-dimensional XY model \cite{Osterloh,AmicoVedral}
and in the antiferromagnetic XXZ model \cite{CaiZhouGuo,GuTianLin1,GuTianLin2}.
Quantum phase transition in spin systems have been widely
characterized considering the pairwise entanglement
between two spin sites \cite{OsborneNielsen}.
The concurrence which is a good measure of entanglement 
\cite{OConnorWooters,Wooters} is maximal close to the critical points
and its derivatives can signal more precisely the presence of a quantum 
phase transition at the critical points \cite{Osterloh}.

In another register Ollivier and Zurek introduced the \emph{quantum discord} 
as a measure of quantum correlations between two separated sub-systems of
a many-body quantum system \cite{OllivierZurek}.
In information theory the \emph{mutual information} is defined as the measure
of correlation between two random variables $\mathcal{A}$ and $\mathcal{B}$.
For classical system the mutual information can be expressed in terms
of two equivalent expressions $\mathcal{I}(\mathcal{A};\mathcal{B})$
and $\mathcal{J}(\mathcal{A};\mathcal{B})$ (explicit definitions of each
expressions of the mutual information are provided later in the text).
However for quantum systems the two expressions are not equal and their difference 
gives the quantum discord. In parallel to the concurrence, quantum discord
measures the quantum correlations of a quantum system. In contrary to the
concurrence, the quantum discord can be different to zero even if a
quantum systems is separable. 
The Werner state, $\rho_{Werner} = \frac{1-\lambda}{4} \UnitMatrix
+ \lambda |\Psi \rangle \langle \Psi|$ where $\lambda$ is a parameter and 
$|\Psi\rangle = \left( |00\rangle + |11\rangle\right)/\sqrt{2}$, 
provide an example for which the concurrence is equal to zero (separable state) 
for $\lambda < 1/3$ while the quantum discord is different to zero 
\cite{OllivierZurek}. Hence quantum discord shows the existence 
of quantum correlations where concurrence shows that the system 
is separable. We point that the concurrence measures the \emph{nonlocal}
\emph{quantumness} of correlations while quantum discord measures 
the total \emph{amount} of quantum correlations of a state $\rho$.

Quantum discord is used in several context \cite{Zurek,RDLutz} for its
property to discernate quantum and classical correlations of a quantum systems.
The behaviour of the quantum as well as the classical correlations close 
to the critical points of QPTs can be studied. 
We focus our attention here on the quantum phase transitions
of the one-dimensional Ising model which derives from the XY model and
on the antiferromagnetic XXZ chain. Using the quantum discord we show that 
the \emph{amount} of quantum correlations increases in the region close
to the critical points. In opposite the \emph{amount} of classical correlations
close to the critical points decreases in the XXZ model while it is simply
monotonous in the Ising model.

The outline of the paper is the following.
In section \ref{Section2} we recall the elements of information 
theory which leads to the construction of quantum discord as 
demonstrated by Ollivier and Zurek \cite{OllivierZurek}.
In section \ref{Section3} and \ref{Section4} the quantum correlations
close to the quantum phase transition of the Ising and XXZ models
are studied. Appendix \ref{AppendixA} provides the derivation of the
quantum discord for spin $S=1/2$ chains.

\section{Quantum discord \label{Section2}}

Information on the correlations between two random variables
$\mathcal{A}$ and $\mathcal{B}$ can be obtained by measure of 
the \emph{mutual information}. In classical information theory 
the mutual information reads

\begin{eqnarray}
\mathcal{I}\left(\mathcal{A} ; \mathcal{B} \right) =
H(\mathcal{A}) + H(\mathcal{B}) 
- H(\mathcal{A},\mathcal{B}),
\label{Eq1}
\end{eqnarray}

\noindent
where the information entropy $H(\mathcal{A}) = - \sum_{a} 
p_{\mathcal{A}=a} \log p_{\mathcal{A}=a}$ and $p_{\mathcal{A}=a}$ is the
probability that $a$ is the realisation of the random variable $\mathcal{A}$.
The joint entropy of $\mathcal{A}$ and $\mathcal{B}$ is defined by
$H(\mathcal{A},\mathcal{B}) = -\sum_{a,b} p_{\mathcal{A}=a,\mathcal{B}=b}
\log p_{\mathcal{A}=a,\mathcal{B}=b}$ where $p_{\mathcal{A}=a,\mathcal{B}=b}$ 
is the joint probability of the variables $\mathcal{A}$ and $\mathcal{B}$
to take respectively the values $a$ and $b$.
Using the Bayes rule $p_{\mathcal{A},\mathcal{B}=b} = p_{\mathcal{B}=b}
\times p_{\mathcal{A}|\mathcal{B}=b}$ the mutual information can be 
rewritten into the equivalent expression

\begin{eqnarray}
\mathcal{J}( \mathcal{A} ; \mathcal{B}) 
= H(\mathcal{A}) - H(\mathcal{A}|\mathcal{B}),
\label{Eq2}
\end{eqnarray}

\noindent
where $H(\mathcal{A}|\mathcal{B}) = -\sum_{a,b} p_{\mathcal{A}=a|\mathcal{B}=b}
\log p_{\mathcal{A}=a|\mathcal{B}=b}$ is the conditional entropy of the random 
variables $\mathcal{A}$ and $\mathcal{B}$ and
$p_{\mathcal{A}=a|\mathcal{B}=b}$ is the conditional probability of $a$ being
the realisation of the random variable $\mathcal{A}$ knowing that $b$ is the
realisation of the random variable $\mathcal{B}$.
For classical random variables the two expressions \eqref{Eq1} and \eqref{Eq2}
are strictly equivalent. However for quantum systems the two expressions
of the mutual information are in general not equivalent
and leads to define the \emph{quantum discord} 
$\delta (\mathcal{A} : \mathcal {B})
= \mathcal{I}(\mathcal{A} ; \mathcal{B}) 
- \mathcal{J}(\mathcal{A} ; \mathcal{B})$ 
as demonstrated in refs. \cite{OllivierZurek,Zurek}.
Quantum discord is the difference between the two classicaly equivalent
expressions of the mutual information.

The mutual information expressions \eqref{Eq1} and \eqref{Eq2} for quantum systems
are obtained by replacing the Shannon entropy by the von Neumann entropy. 
The von Neumann entropy for the joint state reads $H(\mathcal{A},\mathcal{B}) 
= - \tr_{\mathcal{A},\mathcal{B}} \rho_{\mathcal{A},\mathcal{B}} \ln 
\rho_{\mathcal{A},\mathcal{B}}$ where the trace runs over the Hilbert spaces of the 
sub-systems $\mathcal{A}$ and $\mathcal{B}$, $\rho_{\mathcal{A},\mathcal{B}}$
is the density matrix of the quantum system. In a similar way the entropy of the 
system $\mathcal{A}$ reads $H(\mathcal{A}) = - \tr_{\mathcal{A}} \rho_{\mathcal{A}} 
\log \rho_{\mathcal{A}}$ where $\rho_{\mathcal{A}}$ is the reduced density matrix 
obtained by taking the trace over all states of the system $\mathcal{B}$, 
$\rho_{\mathcal{A}} = \tr_{\mathcal{B}} \rho_{\mathcal{A},\mathcal{B}}$.
The quantum conditional entropy $H(\mathcal{A}|\mathcal{B})$ 
in equation \eqref{Eq2} quantifies the ignorance of the state of 
$\mathcal{A}$ knowing the state of $\mathcal{B}$. It is
worked out from the density matrix $\rho_{\mathcal{A} | \Pi_j^{\mathcal{B}}}$
which is related to the density matrix of the system 
$\rho_{\mathcal{A},\mathcal{B}}$ through \cite{OllivierZurek}
\begin{eqnarray*}
\rho_{\mathcal{A} | \Pi_j^{\mathcal{B}}}
=
\Pi_j^{\mathcal{B}} \rho_{\mathcal{A},\mathcal{B}} \Pi_j^{\mathcal{B}} / p_j,
\end{eqnarray*}

\noindent
where $\Pi_j^{\mathcal{B}}$ is the projector onto the state $j$ 
of the sub-system $\mathcal{B}$ and the probability 
$p_j = \tr_{\mathcal{A},\mathcal{B}} \Pi_j^{\mathcal{B}}
\rho_{\mathcal{A},\mathcal{B}}$.
The quantum conditional entropy then reads 
$H(\mathcal{A}|\mathcal{B})= H(\mathcal{A}|\left\{\Pi_j^\mathcal{B}\right\})
=\sum_j p_j H(\rho_{\mathcal{A} | \Pi_j^{\mathcal{B}}})$.

Finaly the quantum discord measuring the quantum correlations between the
two quantum sub-systems $\mathcal{A}$ and $\mathcal{B}$ reads

\begin{eqnarray*}
\delta(\mathcal{A} : \mathcal{B}) =
\min_{ \left\{\Pi_j^\mathcal{B}\right\} }
\Big\{
H(\mathcal{A}) - H(\mathcal{A},\mathcal{B})
+ H(\mathcal{A}|\left\{\Pi_j^\mathcal{B}\right\})
\Big\}.
\end{eqnarray*}

\noindent
It must be noticed that the quantum discord is minimized over
the set of state $\left\{\Pi_j^\mathcal{B}\right\}$.
The information obtained on the sub-system $\mathcal{A}$ 
is affected by the measure on the state of the sub-system $\mathcal{B}$.
In order to get the maximum information on the sub-system $\mathcal{A}$ 
we need to consider the projection onto the states of $\mathcal{B}$ that 
disturb least the overall quantum system. Maximize the information on
the sub-system $\mathcal{A}$ correspond to minimize the quantum discord 
with respects to the set of projector $\left\{\Pi_j^\mathcal{B}\right\}$.

Quantum discord provide information on the quantum nature of the correlations
between two systems. If two sub-systems $\mathcal{A}$ and $\mathcal{B}$ are 
correlated \emph{classicaly} the quantum discord is equal to zero.
Moreover quantum discord shows that quantum correlations can be present
for states that are not entangled. For example in Werner states 
the quantum correlations are still present while the system is separable 
\cite{OllivierZurek}. 
To be more precise we can consider a quantum mixed state 
$\widetilde{\rho} = |\psi\rangle \langle \psi|$ 
acting in an Hilbert space $\mathcal{H}_{\mathcal{A}} \otimes 
\mathcal{H}_{\mathcal{B}}$. The quantum mixed state is separable if its 
entanglement is equal to zero. In this case the quantum separable
state can be written as a product of quantum states and reads
$\widetilde{\rho} = \sum_i p_i |\psi_{i,\mathcal{A}} \rangle 
\langle \psi_{i,\mathcal{A}}| 
\otimes |\psi_{i,\mathcal{B}} \rangle \langle \psi_{i,\mathcal{B}}|$ where
$|\psi_{i,\mathcal{A}} \rangle \in \mathcal{H}_{\mathcal{A}}$,
$|\psi_{i,\mathcal{B}} \rangle \in \mathcal{H}_{\mathcal{B}}$ and 
$\sum_i p_i =1$. The two systems $\mathcal{A}$ and $\mathcal{B}$ can be
correlated through the separable state $\widetilde{\rho}$ and the quantum 
states $|\psi_{i,\mathcal{A}} \rangle$ and $|\psi_{i,\mathcal{B}} \rangle$ 
do not have in general any classical counterpart.
A consequence of the quantumness of the states $|\psi_{\mathcal{A}}\rangle$ and 
$|\psi_{\mathcal{B}}\rangle$ is that the correlations between the two
systems $\mathcal{A}$ and $\mathcal{B}$ have quantum as well as classical
nature. The quantum discord is non-zero while the entanglement
of the quantum separable state is equal to zero.
A good example is furnished by the two-qubit separable mixed state \cite{Datta}

\begin{eqnarray}
\widetilde{\rho} &=& \frac{1}{4} \Bigg(
|\psi_{+}\rangle \langle \psi_{+}|_{\mathcal{A}} 
\otimes |0\rangle \langle 0|_{\mathcal{B}}
+
|\psi_{-}\rangle \langle \psi_{-}|_{\mathcal{A}} 
\otimes |1\rangle \langle 1|_{\mathcal{B}}
\notag \\
&&
+
|0\rangle \langle 0|_{\mathcal{A}}
\otimes |\psi_{-}\rangle \langle \psi_{-}|_{\mathcal{B}}
+
|1\rangle \langle 1|_{\mathcal{A}}
\otimes |\psi_{+}\rangle \langle \psi_{+}|_{\mathcal{B}}
\Bigg),
\notag \\
\label{EqQubit}
\end{eqnarray}

\noindent
where $|\psi_{\pm}\rangle = \frac{1}{\sqrt{2}} 
\left(|0\rangle \pm |1\rangle \right)$ and each qubit $\mathcal{A}$ and
$\mathcal{B}$ have four nonorthogonal states. The two-qubit separable state 
\eqref{EqQubit} show a product of quantum states that can not have any equivalent
classical system and also present quantum correlation. Quantum discord quantifies
all qauntum correlations including entanglement between $\mathcal{A}$ and 
$\mathcal{B}$.

Note that the mutual information $\mathcal{I}(\mathcal{A};\mathcal{B})$ is the
sum of the quantum discord and the classical correlation 
$CC(\mathcal{A};\mathcal{B})$ defined in ref. \cite{HendersonVedral}, 
$\mathcal{I}(\mathcal{A};\mathcal{B}) = \delta(\mathcal{A}:\mathcal{B})
+ CC(\mathcal{A};\mathcal{B})$. In other words classical correlations
$CC(\mathcal{A};\mathcal{B})$ are equal to the mutual information
$\mathcal{J}(\mathcal{A};\mathcal{B})$.

Using quantum discord to discernate the quantum from the classical correlations 
we can study the behaviour of such different correlations in 
quantum phase transitions. In the next sections we focus our attention 
on the quantum phase transitions of the Ising and the XXZ spin-$1/2$ chains
and we study them using the quantum discord.

\section{Quantum discord and quantum phase transition in the Ising chain
\label{Section3}}

The Hamiltonian of the transverse one-dimensionnal Ising model reads

\begin{eqnarray*}
H_{Ising} = - \sum_{i=1}^N 
\left( \lambda \sigma_i^x \sigma_{i+1}^x + \sigma_i^z \right),
\end{eqnarray*}

\noindent
with the boundary condition $\sigma^x_N = \sigma^x_1$ and
$\sigma^\alpha$ with $\alpha = \left\{ x,y,z \right\}$ are the Pauli
matrices and $\sigma^{0} = \UnitMatrix$.
We denotes by $|g\rangle = |\uparrow \rangle$ and $|e\rangle = |\downarrow \rangle$ 
the spin up and spin down states.
For $\lambda = 0$ all spins are pointing in the $z$ direction 
while for $\lambda \rightarrow \infty$ they point in the $x$ direction.
In the thermodynamic limit $N \rightarrow \infty$ the Ising spin chain
undergoes a quantum phase transition at the critical point $\lambda_c = 1$.
The correlation length diverges at this point \cite{OsborneNielsen,CaiZhouGuo}.

The quantum discord is worked out from the joint state of the two spins 
at the lattice sites $i$ and $j$. The information on the joint state is
contained in the two-site density matrix $\rho_{ij}$ which
is derived from the following operator expansion
\begin{eqnarray*}
\rho_{ij} = \tr_{\bar{ij}} \left[ \rho \right] 
= \frac{1}{4} \sum_{\alpha,\beta=0}^3
\Theta_{\alpha \beta} \sigma_i^\alpha \otimes \sigma_j^\beta,
\end{eqnarray*}

\noindent
where the coefficients $\Theta_{\alpha \beta}$ of the expansion are 
related to the spin-spin correlation functions through the relation

\begin{eqnarray*}
\Theta_{\alpha \beta} 
= \tr \left[\sigma_i^\alpha \sigma_j^\beta \rho_{ij} \right]
= \langle \sigma_i^\alpha \sigma_j^\beta  \rangle.
\end{eqnarray*}

\noindent
Owing to the symmetry of the Hamitlonian most of the coefficients 
$\Theta_{\alpha \beta}$ are equal to zero. Translation invariance require that
the density matrix verifies $\rho_{ij}=\rho_{i,i+r}$ and
is independent of the position $i$. 
The reflexion symmetry leads to $\rho_{ij} = \rho_{ji}$,
the Hamiltonian being real the density matrix verifies $\rho_{ij}^{*} =
\rho_{ij}$ and the global rotation symmetry implies that the density matrix
commutes with the operator $\sigma^z_i \sigma^z_j$. Combining all symmetry
constraints the density matrix expressed in the natural
basis $\left\{|gg\rangle, |ge\rangle, |eg \rangle, |ee\rangle \right\}$ 
reduces to \cite{CaiZhouGuo}

\begin{eqnarray}
\rho_{ij}
= \left(
\begin{array}{cccc}
u_{+} & 0 & 0 & y \\
0 & w & x & 0 \\
0 & x & w & 0 \\
y & 0 & 0 & u_{-}
\end{array}
\right),
\label{Eq3}
\end{eqnarray}

\noindent
with 
$
u_{\pm} = \frac{1}{4} \pm \frac{\langle \sigma^z \rangle}{2}
+ \frac{\langle \sigma^z_i \sigma_j^z\rangle}{4}
$,
$w = \frac{1-\langle \sigma^z_i \sigma^z_j \rangle}{4}$,
$x = \frac{\langle \sigma_i^x \sigma^x_j\rangle 
+ \langle \sigma_i^y \sigma^y_j\rangle}{4}$
and
$y = \frac{\langle \sigma_i^x \sigma^x_j\rangle 
- \langle \sigma_i^y \sigma^y_j\rangle}{4}$.
The magnetization of the spin-$1/2$ Ising chain is given by \cite{Barouch1}

\begin{eqnarray*}
\langle \sigma^z \rangle = -\frac{1}{\pi} \int_0^\pi d\phi
\frac{\left(1 + \lambda \cos \phi \right)}{\omega_{\phi}},
\end{eqnarray*}

\noindent
and $\omega_\phi = \sqrt{ (\lambda \sin \phi)^2 + (1+\lambda \cos \phi)^2}$
is the energy spectrum of the transverse Ising chain.
The spin-spin correlations functions are related 
to the determinant of Toeplitz matrices \cite{Barouch2}

\begin{eqnarray*}
\langle \sigma^x_i \sigma^x_{i+r} \rangle &=&
\left|
\begin{array}{cccc}
G_{-1} & G_{-2} & \dots & G_{-r} \\
G_{0} & G_{-1} & \dots & G_{-r+1} \\
\vdots & \vdots & \ddots & \vdots \\
G_{r-2} & G_{r-3} & \dots & G_{-1}
\end{array}
\right|,
 \\
\langle \sigma^y_i \sigma^y_{i+r} \rangle &=&
\left|
\begin{array}{cccc}
G_{1} & G_{0} & \dots & G_{-r+2} \\
G_{2} & G_{1} & \dots & G_{-r+3} \\
\vdots & \vdots & \ddots & \vdots \\
G_{r} & G_{r-1} & \dots & G_{1}
\end{array}
\right|,
\end{eqnarray*}

\noindent
and $\langle \sigma^z_i \sigma^z_{i+r}\rangle = \langle \sigma^z\rangle^2
- G_r G_{-r}$ where

\begin{eqnarray*}
G_{\kappa} &=&
\frac{1}{\pi} \int_0^\pi d\phi \cos(\phi \kappa) 
\frac{(1+\lambda \cos \phi)}{\omega_\phi}
 \\
&&
-\frac{\lambda}{\pi} \int_0^\pi d\phi \sin(\phi \kappa)
\frac{\sin \phi}{\omega_\phi}.
\end{eqnarray*}

The quantum phase transition can be signaled by measure of entanglement.
Indeed the nearest and next-nearest neighbours entanglement reach their maximum 
at the critical point $\lambda_c$ \cite{Osterloh,OsborneNielsen}.
A good measure of the entanglement is provided by the concurrence $\mathcal{C}$.
The concurrence of two spins may be computed from the joint state $\rho_{ij}$
through the formula $\mathcal{C} = \max \left\{ 0, \gamma_1 - \gamma_2 
-\gamma_3 -\gamma_4 \right\}$ where the $\gamma_i$ are the eigenvalues
in decreasing order of the matrix $R = \sqrt{\rho_{ij} \widetilde{\rho_{ij}}}$
 \cite{OConnorWooters,Wooters}.
The matrix $\widetilde{\rho_{ij}}$ is related to the transpose of the 
two-site density matrix by $\widetilde{\rho_{ij}} = \left( \sigma^y \otimes
\sigma^y\right) \rho_{ij}^{*} \left( \sigma^y \otimes \sigma^y\right)$.
Hence the entanglement between two spins at site $i$ and $j$ is given by
\begin{eqnarray*}
\mathcal{C}_{ij} = 2 \max \left\{ 0, |x| - \sqrt{u_{+} u_{-}}, |y| - w \right\}.
\end{eqnarray*}

\begin{figure}
\center
\epsfig{file=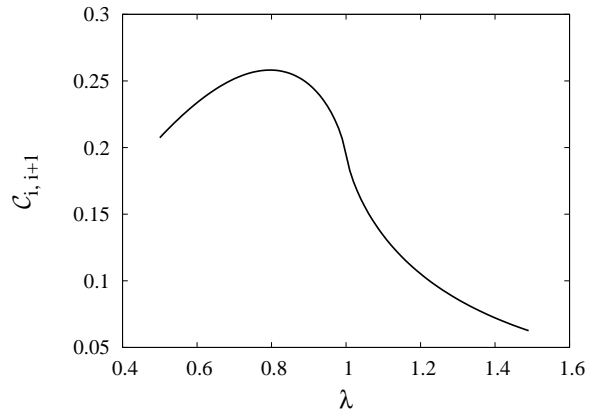,width=8cm}
\caption{Concurrence for nearest neighbour spins in the Ising spin $S=1/2$
chain. The concurrence is maximal close to the critical coupling $\lambda_c = 1$.}
\label{Fig1}
\end{figure}

\begin{figure}
\center
\epsfig{file=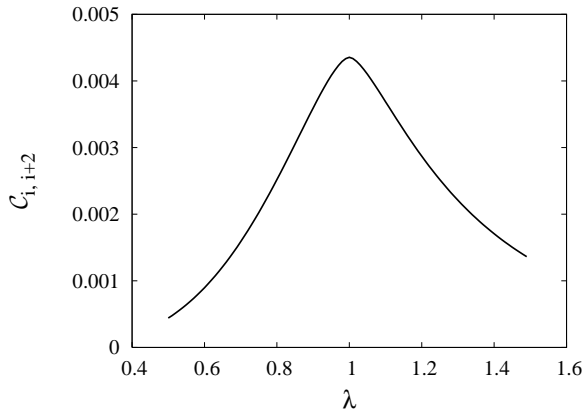,width=8cm}
\caption{Concurrence for next-nearest neighbour spins in the Ising spin chain. 
The concurrence is maximal at the critical coupling $\lambda_c = 1$.}
\label{Fig1bis}
\end{figure}

\noindent
Figures \ref{Fig1} and \ref{Fig1bis} represent respectively the concurrence for the 
nearest-neighbour spins $\mathcal{C}_{i,i+1}$ and next-nearest neighbour spins 
$\mathcal{C}_{i,i+2}$. It shows that the maximum of entanglement is reached
close to $\lambda_c = 1$. Only the derivatives of the concurrence show
singularities at the critical points \cite{Osterloh} signaling the presence
of the quantum phase transition.

Entanglement increases close to the quantum phase transition. However entanglement
is only one kind of quantum correlation. It is then legitimate to ask :
What is the behaviour of the \emph{total amount} of quantum correlations close to 
the critical point of the QPT ? The answer is provided by the quantum discord.
The details of the derivation of the quantum discord are given
in appendix \ref{AppendixA}. Figure \ref{Fig2} represents the quantum
discord for the Ising $S=1/2$ chain for nearest-neighbour $\delta_{i,i+1}$
and next-nearest-neighbour $\delta_{i,i+2}$. It shows that the quantum
correlations increase and are maximal in a region close to the critical
point $\lambda_c$.

\begin{figure}
\center
\epsfig{file=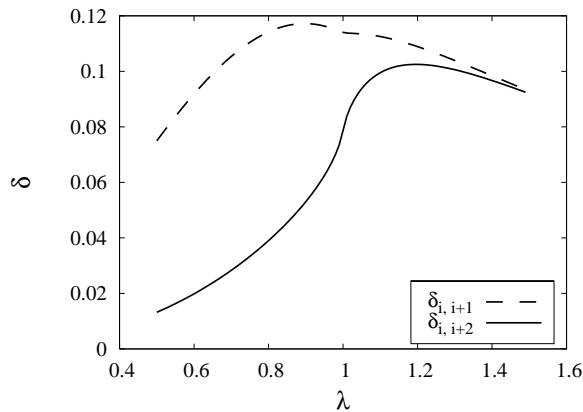,width=8cm}
\caption{Quantum discord $\delta$ for the nearest-neighbour (in dashed line) and 
next-nearest-neighbour (in full line) spin sites of the one-dimensionnal 
Ising model. The quantum discord is minimized for $\phi=0$ and $\theta$ 
varying progressively from zero for $\lambda<1$ to $\pi/4$ for $\lambda > 1$, 
see appendix \ref{AppendixA} for the details on the derivation.}
\label{Fig2}
\end{figure}

\begin{figure}
\center
\epsfig{file=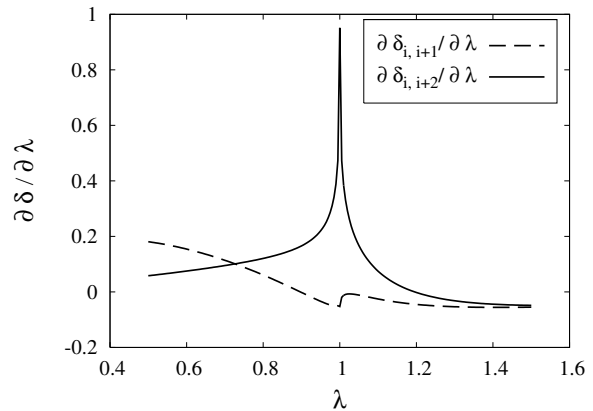,width=8cm}
\caption{First derivatives of the quantum discord with respects to the coupling
parameter $\lambda$ for the nearest-neighbour (in dashed line) and 
next-nearest-neighbour (in full line) spin sites of the one-dimensionnal 
Ising model.}
\label{Fig2bis}
\end{figure}

More informations on the location and the order of the quantum phase transition
can be obtained by consideration of the derivatives of the quantum discord with
respects to the coupling parameter $\lambda$. Indeed there is a relation between
the nonanalyticity in energy and quantum discord that leads to the characterization
of the quantum phase transition \cite{WuSarandyLidar}.
Consider the energy of the spin system in terms of the reduced 
density matrix of two spins at positions $i$ and $j$, the energy reads

\begin{eqnarray}
E\left( \rho_{ij} \right) = \sum_{ij} 
\tr \left[\widetilde{H}_{ij} \rho_{ij} \right],
\label{EqEnergy}
\end{eqnarray}

\noindent
where $\widetilde{H}_{ij}$ is the reduced Hamiltonian of the two spins 
at position $i$ and $j$. The sum of $\widetilde{H}_{ij}$ over the positions $i,j$
is equal to the Hamiltonian of the spin system, 
$\sum_{ij} \widetilde{H}_{ij} \equiv H$.
For the one-dimensionnal Ising model the reduced Hamiltonian reads 
$\widetilde{H}_{ij}^{Ising} = - \left(\lambda \sigma_i^x \sigma_{j}^x 
\widetilde{\delta}_{i+1,j} + \sigma_i^z/N \right)$ where $N$ is the number 
of spins and $\widetilde{\delta}_{ij}$ is the Kronecker delta. 
The order as well as the location of a quantum phase
transition can be characterized by the nonanaliticity of the energy.
If the first derivative of the ground state energy presents a finite discontinuity
then the quantum phase transition is of the first order. However if the first
derivative is continuous while the second derivative shows discontinuity or divergence
then the quantum phase transition is of the second order. Moreover the derivatives
of the energy are related to the derivatives of the reduced density matrix and
one can show that \cite{WuSarandyLidar}
\begin{eqnarray*}
\frac{\partial E \left( \rho_{ij} \right)}{\partial \lambda} &=& 
\sum_{ij} \tr \left[ \frac{\partial \widetilde{H}_{ij}}{\partial \lambda} 
\rho_{ij} \right],
\notag \\
\frac{\partial^2 E \left( \rho_{ij} \right)}{\partial \lambda^2} &=& 
\sum_{ij} \Bigg\{
\tr \left[ \frac{\partial^2 \widetilde{H}_{ij}}{\partial \lambda^2} 
\rho_{ij}\right]
+
\tr \left[ \frac{\partial \widetilde{H}_{ij}}{\partial \lambda} 
\frac{\partial \rho_{ij}}{\partial \lambda} \right]
\Bigg\}.
\end{eqnarray*}

\noindent
Note that the derivatives of the reduced Hamiltonian are continuous with respects
to the coupling parameter $\lambda$. Hence a discontinuity in the first derivative
$\frac{\partial E \left( \rho_{ij} \right)}{\partial \lambda}$
of the energy at the critical point implies a discontinuity at least of one of the 
reduced density matrices $\rho_{ij}$. Similarly a discontinuity or a singularity in 
the second derivatives of the energy 
$\frac{\partial^2 E \left( \rho_{ij} \right)}{\partial \lambda^2}$ requires
the divergence of at least one of the derivatives 
$\frac{\partial \rho_{ij}}{\partial \lambda}$ at the critical points.
It becomes now evident that the quantum phase transition can be characterized
by an analysis of the derivatives of the quantum discord with respects to the
coupling parameter $\lambda$. The quantum discord being dependent on the reduced
density matrix one deduces that i) a discontinuity in the quantum discord implies
a discontinuity in $\rho_{ij}$ (hence the first derivative of the energy is 
discontinuous) and a first order QPT, ii) a singularity in the derivative of the 
quantum discord implies a discontinuity or a divergence of 
$\frac{\partial \rho_{ij}}{\partial \lambda}$ (the second derivative of the energy
is discontinuous) and the QPT is of the second order.

Figure \ref{Fig2bis} shows that for nearest neighbours 
$\partial \delta_{i,i+1}/ \partial \lambda$ presents a discontinuity at the 
critical point and for the next-nearest neighbours 
$\partial \delta_{i,i+2}/ \partial \lambda$ shows a singularity. 
The fact that the quantum discord is continuous while its derivatives are 
discontinuous or singular at $\lambda=\lambda_c$ indicates that the precise 
location of the critical point is at $\lambda_c=1$ and that the QPT is of 
the second order.

\begin{figure}
\center
\epsfig{file=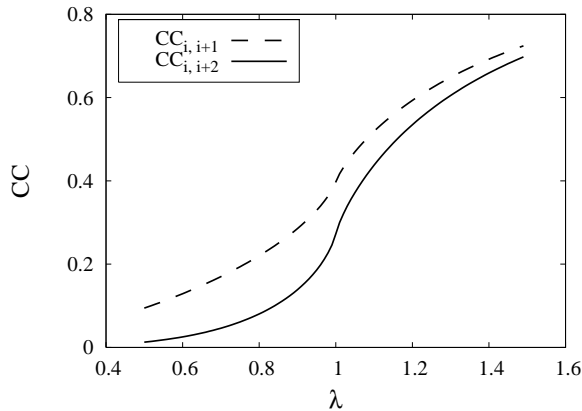,width=8cm}
\caption{Classical correlation $CC$ for the nearest-neighbour (in dashed line) and 
next-nearest-neighbour (in full line) spin sites of the one-dimensionnal 
Ising model.}
\label{Fig3}
\end{figure}

\begin{figure}
\center
\epsfig{file=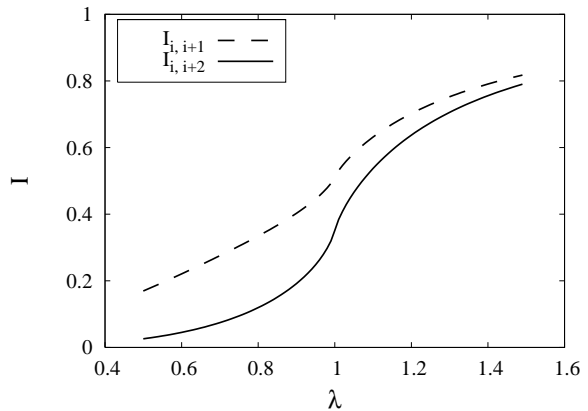,width=8cm}
\caption{Mutual information $\mathcal{I}(\mathcal{A};\mathcal{B})$
for the nearest-neighbour (in dashed line) and next-nearest-neighbour
(in full line) spin sites of the one-dimensionnal Ising model.}
\label{Fig4}
\end{figure}

For $\lambda > 1$ and approching the critical point from the right, 
the classical correlations decrease in opposite to the quantum discord 
which increases as depicted by the figures \ref{Fig2} and \ref{Fig3}. 
For $\lambda < 1$ both classical and quantum correlations are decreasing 
when the coupling parameter $\lambda$ decreases. Classical correlations 
monotonically increase with respects to $\lambda$ while quantum discord 
show that the amplitude of the quantum correlations increase close to the 
critical point $\lambda_c$.
From the mutual information we infer that the spins are uncorrelated for 
$\lambda \rightarrow 0$ as depicted in figure \ref{Fig4} and can be confirmed 
by the Hamiltonian which reduces to $H_{Ising,\lambda = 0} = -\sum_{i=1}^N \sigma_j^z$.
Close to the critical point the \emph{amount} of quantum correlations is maximal
as supported by the concurrence.

\section{Quantum discord and quantum phase transition in the XXZ chain
\label{Section4}}

The hamiltonian of the one-dimensionnal XXZ spin-$1/2$ chain reads

\begin{eqnarray*}
H_{XXZ} = \sum_{\langle ij \rangle} \left(S_i^x S_j^x + S_i^y S_j^y 
+ \Delta S_i^z S_j^z \right),
\end{eqnarray*}

\noindent
where the sum runs over the nearest-neighbour bonds $\langle i j \rangle$.
At the critical point $\Delta_c = 1$ the XXZ spin chain undergoes
a quantum phase transition between an XY phase for $-1 < \Delta < 1$ 
and an Ising antiferromagnetic phase for $\Delta > 1$.

The density matrix of two nearest-neighbour spin sites is given by \eqref{Eq3}.
However owing to the spin-flip symmetry the magnetization is equal to zero, 
$\langle \sigma^z \rangle =0$ and the XXZ Hamiltonian is also symmetric with 
respects to the rotation in the $xy$ plan which leads to the equality between
the spin-spin correlation function in the $xy$ plan 
$\langle \sigma_i^x \sigma_j^x \rangle = \langle \sigma_i^y \sigma_j^y \rangle$. 
Gathering all the symmetry constraints of the XXZ Hamiltonian the joint density 
matrix $\rho_{ij}$ reduces to
\begin{eqnarray*}
\rho_{ij} = 
\left(
\begin{array}{cccc}
u & 0 & 0 & 0 \\
0 & w & x & 0 \\
0 & x & w & 0 \\
0 & 0 & 0 & u
\end{array}
\right),
\label{Eq4}
\end{eqnarray*}

\noindent
with 
$
u_{\pm} = u = \frac{1}{4}
+ \frac{\langle \sigma^z_i \sigma_j^z\rangle}{4}
$,
$w = \frac{1-\langle \sigma^z_i \sigma^z_j \rangle}{4}$,
$x = \frac{\langle \sigma_i^x \sigma^x_j\rangle }{2}$
and
$y = 0$.

The spin-spin correlation functions between nearest-neighbour spin sites
for $-1 < \Delta < 1$ are given by \cite{Jimbo1,Korepin,Kato}

\begin{eqnarray*}
\langle \sigma_i^x \sigma_{i+1}^x \rangle &=&
\frac{\cos \pi \Phi}{\pi^2} \int_{-\infty}^{\infty} \frac{dx}{\sinh x}
\frac{x \cosh x}{\cosh^2 (\Phi x)}
\notag \\
&&
-
\frac{1}{\pi \sinh (\pi \Phi)} \int_{-\infty}^{\infty} \frac{dx}{\sinh x}
\frac{\sinh \left((1-\Phi)x\right)}{\cosh \Phi x},
\end{eqnarray*}

\noindent
where $\Delta = \cos \left(\pi \Phi \right)$, and

\begin{eqnarray*}
\langle \sigma_i^z \sigma_{i+1}^z \rangle &=&
1 - \frac{2}{\pi^2} \int_{-\infty}^{\infty} \frac{dx}{\sinh x}
\frac{x \cosh x}{\cosh^2 (\Phi x)}
\notag \\
&&
+ \frac{2 \cot (\pi \Phi)}{\pi} \int_{-\infty}^{\infty} \frac{dx}{\sinh x}
\frac{\sinh \left((1-\Phi)x\right)}{\cosh \Phi x}.
\end{eqnarray*}

\noindent
And for $\Delta > 1$ the correlation function are given by 
\cite{Jimbo2,Nakayashiki,Takahashi}

\begin{eqnarray*}
\langle \sigma_i^x \sigma_{i+1}^x \rangle &=&
\int_{-\infty + i/2}^{\infty + i/2} \frac{dx}{\sinh (\pi x)}
\notag \\
&& \times
\left(
\frac{x}{\sin^2 (\phi x)}\cosh \nu
- \frac{\cot (\nu x)}{\sinh \nu}
\right),
\end{eqnarray*}

\noindent
with $\Delta = \cosh \nu$, and

\begin{eqnarray*}
\langle \sigma_i^z \sigma_{i+1}^z \rangle &=&
1 + 2 \int_{-\infty + i/2}^{\infty + i/2} \frac{dx}{\sinh (\pi x)}
\notag \\
&& \times
\left(
\cot (\nu x) \coth (\nu) - \frac{x}{\sin^2 (\nu x)}
\right).
\end{eqnarray*}

\begin{figure}
\center
\epsfig{file=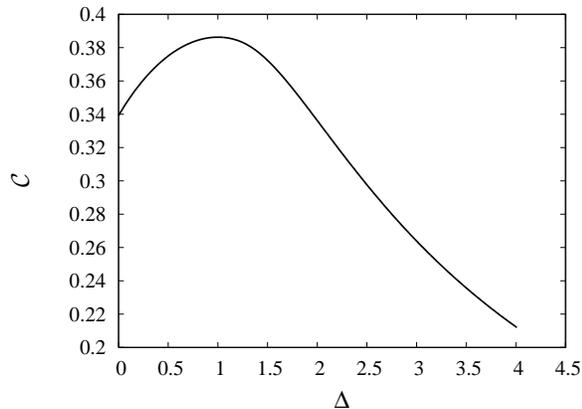,width=8cm}
\caption{Concurrence for nearest-neighbour spin sites in 
the XXZ spin $S=1/2$ one-dimensionnal model.}
\label{Fig5}
\end{figure}

\begin{figure}
\center
\epsfig{file=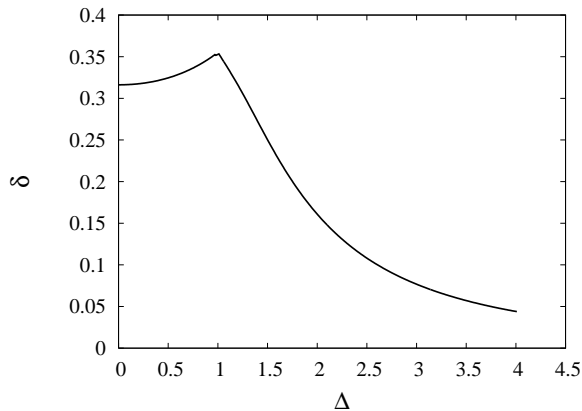,width=8cm}
\caption{Quantum discord for nearest neighbour spins in the XXZ spin $S=1/2$ 
one-dimensionnal model. The quantum discord is minimized for $\phi=0$ and $\theta$ 
varying progressively from zero for $\lambda<1$ to $\pi/4$ for $\lambda > 1$, 
see appendix \ref{AppendixA} for the details on the derivation.}
\label{Fig6}
\end{figure}

\begin{figure}
\center
\epsfig{file=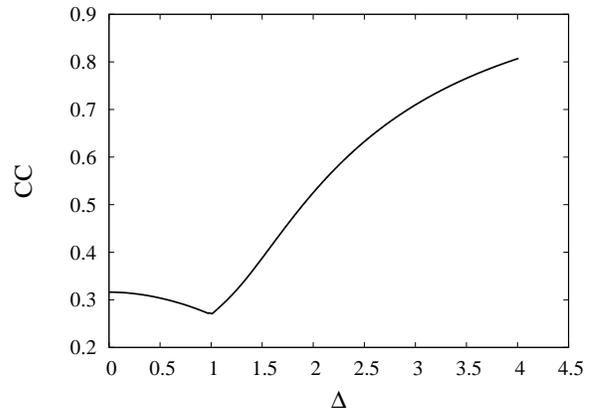,width=8cm}
\caption{
Dependence of the classical correlation 
$CC(\mathcal{A};\mathcal{B})$ on the parameter $\Delta$ 
for the nearest-neighbour spin sites in the XXZ spin $S=1/2$ chain.}
\label{Fig7}
\end{figure}

The concurrence for the XXZ spin-$1/2$ chain is given by
$\mathcal{C} = 2 \max \left\{0,|x|-|u|\right\}$.
Figure \ref{Fig5} represents the dependence of the concurrence with respects
to the control parameter $\Delta$ and shows that entanglement between the 
nearest-neighbour spin sites is maximal at the critical point $\Delta_c = 1$ 
\cite{CaiZhouGuo}.

In parallel the quantum discord presents the maximum of the
quantum correlations at the critical point. The derivation of the
quantum discord for the XXZ model follow the same scheme as for the
Ising model and the details of the derivation are provided in
appendix \ref{AppendixA}. Figure \ref{Fig6} depicts the dependence
of the quantum discord on the control parameter $\Delta$. The quantum
discord is minimized with respects to the angles $\phi$ and $\theta$
that control the axes of the projector $\Pi_j^{\mathcal{B}}$.

As depicted in figure \ref{Fig6} the quantum correlations are
stronger close to the QPT. On the opposite 
the classical correlations are weaker close to the critical point
$\Delta_c$. Both classical and quantum correlations measured by
quantum discord and the mutual information 
$\mathcal{J}(\mathcal{A};\mathcal{B})$ behave accordingly to the 
behaviour of entanglement measured by the concurrence.

The cusps observed in figures \ref{Fig6} and \ref{Fig7} for both the classical 
correlation and the quantum discord arise from a level-crossing between the 
ground-state energy and an excited-state energy. This scheme explains
the cusp observed in the concurrence measuring the entanglement of nearest 
neighbour spin in $(d \ge 2)$-dimensionnal XXZ model \cite{GuTianLin1}. A similar
mechanism explains also the fact that the concurrence is maximal at the critical 
points for the one-dimensionnal XXZ model \cite{GuTianLin3}.
Level crossing induces a qualitative change in the ground states at the 
critical point $\Delta_c = 1$ and leads to the abrupt behaviour of the
quantum discord and classical correlation at the quantum phase transition.
As depicted in figure \ref{Fig5} the concurrence is less sensitive to
the level-crossing than the quantum discord and the classical correlations 
since it does not present an abrupt cusp at $\Delta_c$.

Remark that the quantum discord for the next-nearest neighbour spins 
$\delta_{i,i+2}$ is not presented for the 1d XXZ model. The reason is that no more 
informations can be gained by $\delta_{i,i+2}$ than already furnished 
by the quantum discord for nearest neighbours $\delta_{i,i+1}$. 
Indeed the quantum discord $\delta_{i,i+2}$ presents a similar dependence on 
the coupling parameter $\Delta$ as shown by $\delta_{i,i+1}$ and its amplitude 
verifies $\delta_{i,i+1} \sim 2 \delta_{i,i+2}$.

\section{Conclusion \label{Section5}}

Quantum phase transition of the one-dimensional spin-$1/2$ Ising and
antiferromagnetic XXZ models have been studied using the quantum discord 
\cite{OllivierZurek,Zurek}.
Quantum discord is a measure of the quantum correlation given by
the difference of two classicaly equivalent expressions of the mutual
information. Using quantum discord and mutual information 
we showed that it is possible to discernate the quantum and the classical nature 
of the correlations between two elements of a quantum many-body system.

We demonstrated that the amount of quantum correlations is larger
close to the critical points for the quantum phase transitions of the 
one-dimensionnal XXZ and Ising models. For the XXZ model the amount of 
classical correlations decrease at the critical point while for the Ising
model they are simply monotonous with respects to the coupling parameter 
$\lambda$. The behaviours of the quantum and classical correlations measured
by means of the quantum discord are in agreement with the behaviour of 
the entanglement computed by the concurrence for both spin models.

Quantum discord is a good way to discernate the nature of the correlations
between the components of a quantum system and it is a good qualitative
indicator of the existence of quantum phase transition. 
However depending on the model studied the maximum of the quantum discord 
is not necessarily located at the critical points as shown for the Ising model. 
The quantum discord can only be used for qualitative detection of 
quantum phase transitions. On the contrary the derivatives of the quantum
discord provide precise informations on the location and on the order
of the quantum phase transitions. Quantum discord \cite{OllivierZurek,Zurek} and 
concurrence \cite{GuTianLin1,GuTianLin2,CaiZhouGuo,OsborneNielsen} are two useful
tools to detect quantum phase transitions. Despite the fact that implementation of 
concurrence is more easy/rapid than for quantum discord, concurrence measures only
the nonlocal quantumness of the correlations between two systems $\mathcal{A}$
and $\mathcal{B}$. Quantum discord measures the total amount of quantum correlations
between $\mathcal{A}$ and $\mathcal{B}$ including entanglement.

We showed that quantum discord agrees with the predictions
provided by the concurrence for the behaviour of the quantum 
correlations close to quantum phase transitions for one-dimensionnal
quantum systems. It would be interesting to confirm this observation 
in futur investigation of quantum systems with higher dimensionnality.

\acknowledgments{The author would like to thank Eric Lutz for useful discussions.
This work was supported by  the Emmy Noether Program of the DFG 
(Contract LU1382/1-1) and the cluster of excellence Nanosystems Initiative 
Munich (NIM).}

\appendix
\section{Derivation of the quantum discord for spin chains \label{AppendixA}}

The quantum discord for two systems reads 
$\delta(\mathcal{A} : \mathcal{B}) =
\min_{ \left\{\Pi_j^\mathcal{B}\right\} }
\Big\{
H(\mathcal{A}) - H(\mathcal{A},\mathcal{B})
+ H(\mathcal{A}|\left\{\Pi_j^\mathcal{B}\right\})
\Big\}$ for which the three von Neumann entropy are computed
over respectively the reduced density matrix $\rho_{\mathcal{A}}$,
the joint state $\rho_{\mathcal{A},\mathcal{B}}$ and the conditional
density matrix $\rho_{\mathcal{A}|\Pi_j^{\mathcal{B}}}$.

The reduced density matrix $\rho_{\mathcal{A}}$ is given by
\begin{eqnarray*}
\rho_{\mathcal{A}} &=& \tr_{\mathcal{B}} \rho_{\mathcal{A},\mathcal{B}}
 \\
&=& 
\left( u_{+}+w \right) |g \rangle \langle g |_{\mathcal{A}}
+
\left( u_{-}+w \right) |e \rangle \langle e |_{\mathcal{A}},
\end{eqnarray*}

\noindent
which leads to the von Neumann entropy of the reduced density matrix
\begin{eqnarray}
H(\mathcal{A}) &=& - \tr_{\mathcal{A}} \rho_{\mathcal{A}} \log \rho_{\mathcal{A}}
\notag \\
&=& 
- \frac{1}{2} \left(1 + \langle \sigma^z \rangle \right)
\log \left[ \frac{1}{2} \left(1 + \langle \sigma^z \rangle \right) \right]
\notag \\
&&
- \frac{1}{2} \left(1 - \langle \sigma^z \rangle \right)
\log \left[ \frac{1}{2} \left(1 - \langle \sigma^z \rangle \right) \right].
\label{EqA1}
\end{eqnarray}

The joint density matrix is equal to the density matrix
given in equation \eqref{Eq3} $\rho_{\mathcal{A},\mathcal{B}}=\rho_{ij}$
and the von Neumann entropy is given yb the sum of the Shannon entropy
over the eigenvalues of the density matrix $\rho_{ij}$ and reads
\begin{eqnarray}
H(\mathcal{A},\mathcal{B}) &=& -\tr_{\mathcal{A},\mathcal{B}} 
\rho_{\mathcal{A},\mathcal{B}} \log \rho_{\mathcal{A},\mathcal{B}}
\notag \\
&=&
- \left(w+x \right) \log \left(w+x \right)
- \left(w-x \right) \log \left(w-x \right)
\notag \\
&&
- \sum_{\epsilon = \pm} \Xi_{\epsilon} \log \Xi_{\epsilon},
\label{EqA2}
\end{eqnarray}

\noindent
where 
\begin{eqnarray*}
\Xi_{\pm} = 
\frac{1}{2}
\Big\{
(u_{+}+u_{-}) \pm \left[ (u_{+}-u_{-})^2 + 4 y^2 \right]^{1/2}
\Big\}.
\end{eqnarray*}

The conditional density matrix $\rho_{\mathcal{A}|\Pi_j^{\mathcal{B}}}$
is given by projection on the arbitrary basis
\begin{eqnarray*}
|g\rangle_{\mathcal{B}} &=& \cos(\theta) | j_1 \rangle_{\mathcal{B}}
+ e^{i \phi} \sin(\theta) |j_2 \rangle_{\mathcal{B}},
 \\
|e\rangle_{\mathcal{B}} &=& e^{-i\phi}\sin(\theta) | j_1 \rangle_{\mathcal{B}}
- \cos(\theta) |j_2 \rangle_{\mathcal{B}}.
\end{eqnarray*}

\noindent
The angles $\phi$ and $\theta$ control the projectors direction and the
quantum discord is minimized with repsects to these angles.

\noindent
The conditional entropy is then given by 

\begin{eqnarray*}
\rho_{\mathcal{A}|\Pi_j^{\mathcal{B}}} &=&
\Pi_j^{\mathcal{B}} \rho_{\mathcal{A},\mathcal{B}} \Pi_j^{\mathcal{B}} / p_j ,
\end{eqnarray*}

\noindent
with $\Pi_{j_{\alpha=1,2}}^{\mathcal{B}} = \UnitMatrix_{\mathcal{A}} \otimes 
| j_{\alpha} \rangle \langle j_{\alpha}|_{\mathcal{B}}$ and $p_{j_\alpha} 
= \tr_{\mathcal{A},\mathcal{B}} \UnitMatrix_{\mathcal{A}} \otimes
| j_{\alpha} \rangle \langle j_{\alpha}|_{\mathcal{B}} \rho_{\mathcal{A},\mathcal{B}}$.

For the projection onto the state $|j_\alpha\rangle$ where $\alpha =1,2$ 
one gets

\begin{eqnarray}
\rho_{\mathcal{A}|\Pi_{j_\alpha}^{\mathcal{B}}} &=&
|j_\alpha\rangle \langle j_\alpha|_{\mathcal{B}} \otimes 
\Bigg\{
|g\rangle \langle g|_{\mathcal{A}} 
X_{j_\alpha,+}
+
|e\rangle \langle e|_{\mathcal{A}} 
X_{j_\alpha,-}
\notag \\
&&
+
|g\rangle \langle e|_{\mathcal{A}} 
Y_{j_\alpha}
+
|e\rangle \langle g|_{\mathcal{A}} 
\bar{Y}_{j_\alpha}
\Bigg\}/p_{j_\alpha} ,
\label{EqA3}
\end{eqnarray}

\noindent
where for $\alpha =1$, 

\begin{eqnarray*}
X_{j_1,+} &=& \left(u_{+} \cos^2(\theta) + w \sin^2(\theta) \right),
 \\
X_{j_1,-} &=& \left(u_{-} \sin^2(\theta) + w \cos^2(\theta) \right),
 \\
Y_{j_1} &=& \left(e^{-i\phi}x + y e^{i\phi} \right) \sin(\theta) \cos(\theta),
\end{eqnarray*}

\noindent
and with $p_{j_1} = \frac{1}{2}\left(1 + \langle \sigma^z \rangle 
\cos(2 \theta) \right)$. 
For the projection onto the state $|j_2\rangle$ the elements $X_{j_2,\pm}$ are
given by 

\begin{eqnarray*}
X_{j_2,+} &=& \left(u_{+} \sin^2(\theta) + w \cos^2(\theta) \right),
 \\
X_{j_2,-} &=& \left(u_{-} \cos^2(\theta) + w \sin^2(\theta) \right),
 \\
Y_{j_2} &=& -\left(e^{-i\phi}x + y e^{i\phi} \right) \sin(\theta) \cos(\theta),
\end{eqnarray*}

\noindent
and with $p_{j_2} = \frac{1}{2}\left(1 - \langle \sigma^z \rangle 
\cos(2 \theta) \right)$.
The eigenvalues $\Psi_{j_\alpha,\pm}$ of the 
conditional density matrix \eqref{EqA3} reads
$
\Psi_{j_\alpha,\pm} = 
\frac{1}{2 p_{j_\alpha}} \Bigg\{
(X_{j_\alpha,+} + X_{j_\alpha,-} ) \pm \left[(X_{j_\alpha,+}-X_{j_\alpha,-})^2 
+ 4 Y_{j_\alpha} \bar{Y}_{j_\alpha}  \right]^{1/2}
\Bigg\}
$.

Finaly the conditional von Neumann entropy reads

\begin{eqnarray}
H(\mathcal{A}|\left\{\Pi_j^\mathcal{B}\right\})
&=&
\sum_{j=1,2} - p_j \tr_{\mathcal{A}} \rho_{\mathcal{A}|\Pi_{j}^{\mathcal{B}}}
\log \rho_{\mathcal{A}|\Pi_{j}^{\mathcal{B}}}
\notag \\
&=&
- \sum_{\epsilon=\pm} \sum_{j=1,2} p_j \Psi_{j,\epsilon}(\phi,\theta) 
\log \Psi_{j,\epsilon}(\phi,\theta).
\notag \\
\label{EqA5}
\end{eqnarray}

The quantum discord is minimized with respects to the angles $\phi$ and
$\theta$ that control the direction of the projector $\Pi_j^{\mathcal{B}}$ 
in the Hilbert space of the sub-system $\mathcal{B}$. 
The minimization must be achieved for each ensemble of values
of the magnetization and the spin-spin correlations function which defines
the joint density matrix. It can be shown numericaly that for the 
one-dimensionnal Ising and antiferromagnetic XXZ models
the set of angles $\left\{\phi,\theta \right\}$ that minimize the 
quantum discord belong to the ranges $\phi=0, \theta \in \left[0,\pi/4 \right]$.

\end{document}